\documentclass[draftcls,onecolumn,12pt]{IEEEtran}

\usepackage{graphicx,amsmath,amssymb,array,psfrag,amsbsy,dsfont}
\usepackage{multirow}
\usepackage[noadjust]{cite}
\newtheorem{theorem}{\textbf{Theorem}}
\newtheorem{definition}{\textbf{Definition}}%[section]
\newtheorem{lemma}{\textbf{Lemma}}%[section]
\usepackage{times}
\usepackage{epsf}
\usepackage{epsfig}
\usepackage{amsmath}
\usepackage{amsfonts}
\usepackage{amssymb}
\usepackage{amstext}
\usepackage{latexsym}
\usepackage{color}
\usepackage{ifthen}
\usepackage{multirow}
\usepackage{subfigure}

\usepackage{color,cite,times,amstext,latexsym,ifthen}
\usepackage{graphicx,amsfonts,epsf,mathrsfs}
\usepackage{verbatim}
\usepackage{dsfont}

\usepackage{bbold}

\usepackage{graphicx,epsfig}
\usepackage{mathrsfs}
\usepackage{amsfonts}
\usepackage{color}
\usepackage{fancybox}
\usepackage{amssymb}
\usepackage{amsmath}

\renewcommand{\P}{\mathbf{P}}
\newcommand{\T}{\mathcal{T}}
\renewcommand{\H}{\mathscr{H}}
\newcommand{\cC}{\mathcal{C}}
\newcommand{\cS}{\mathcal{S}}
\newcommand{\cH}{\mathcal{H}}
\newcommand{\cL}{\mathcal{L}}
\newcommand{\cR}{\mathcal{R}}

\newtheorem{remark}{Remark}

\newcommand{\qed}{$\hfill\blacksquare$}

\title{Tight Bounds on the Redundancy of Huffman Codes
\thanks{The material in this paper was presented in part at the IEEE Information Theory Workshop, Punta del Este, Uruguay, March 2006.}
}

\date{}
\begin{document}

\author{Soheil Mohajer, Payam Pakzad, and  Ali Kakhbod
\thanks{Soheil Mohajer was with the School of Computer and Communication Sciences, Ecole Polytechnique F\'{e}d\'{e}rale de Lausanne
(EPFL), CH-1015 Lausanne, Switzerland. He is now with the Department of Electrical Engineering and Computer Sciences,
University of California at Berkeley, Berkeley, CA 94720, USA (email: mohajer@eecs.berkeley.edu). 
Payam Pakzad is with Qualcomm Research,  Silicon Valley,  3165 Kifer Road, Santa Clara, CA 95051, USA (email: payam@qualcomm.com).
Ali Kakhbod is with the Department of Electrical Engineering and Computer Science, University of Michigan, Ann Arbor, MI 48109, USA (email: akakhbod@umich.edu). }}

%\author{Soheil Mohajer,~\IEEEmembership{Student Member,~IEEE}, Chao Tian,~\IEEEmembership{Member,~IEEE},  and\\ Suhas N. Diggavi,~\IEEEmembership{Member,~IEEE}
%\thanks{Soheil Mohajer and Suhas N. Diggavi are with the School of Computer and Communication Sciences, Ecole Polytechnique Federale de Lausanne, Switzerland. Chao Tian  is with the 
%AT\&T Labs-Research, Florham Park, New Jersey, USA.}
%}

\maketitle

\begin{abstract}
In this paper we study the redundancy of Huffman codes. In particular, we consider sources for which the probability of one of the source symbols is known. We prove a conjecture of Ye and Yeung regarding the upper bound on the redundancy of such Huffman codes, which yields in a tight upper bound. We also derive a tight lower bound for the redundancy under the same assumption.

We further apply the method introduced in this paper to other related problems. It is shown that several other previously known bounds with different constraints follow immediately from our results.
\end{abstract}

\begin{IEEEkeywords}
Huffman Code, redundancy, tight bounds.
\end{IEEEkeywords}

\section{Introduction}
Consider a discrete memoryless source $S$ with finite alphabet  $\mathcal{S}=\{s_1,s_2,\dots,s_N\}$ of size $N$, and with the multiset\footnote{$\P$ is a multiset since it may contain repeated members, due to different source symbols with the same probabilities. We will define the notion of multiset and its properties in detail in Definition~\ref{def:multiset}.}  of probabilities $\P=\{u_1, u_2, \dots, u_N\}$, where $u_i=\Pr(S=s_i)$. A variable-length $D$-ary lossless source code for the source $S$ is defined by a one-to-one encoding function 
\begin{align*}
f_{\cC}:\mathcal{S}\longrightarrow \{0,1,\dots,D-1\}^{\star},
\end{align*}
where $A^{\star}=\bigcup_{n\in\mathds{N}} A^n$. Moreover, a code is called \emph{prefix-free} if
there is no valid codeword in the code that is a prefix of any other valid codeword. The length of the codeword assigned to the symbol $s_i$ (with relative frequency $u_i$) by the code $\cC$ is denoted by $l_{\cC}(u_i)$. Therefore, the \emph{average codeword length} of a code $\cC$ is defined as
\begin{align}
\cL(\cC):=\sum_{i=1}^N u_i l_\cC(u_i).
\end{align}
A basic goal in compression is to design codes with minimum average length. It is known that the average length of a prefix-free code cannot be less than the source \emph{entropy} regardless of how efficient the code is, where the entropy  of the source (in base $D$) is defined as
\begin{equation}
\cH(\cS):=-\sum_{i=1}^N u_i \log_D(u_i).
\end{equation}
An important performance measure of a source code, called the redundancy of the code, is the difference of its average length and the entropy of the source, that is, 
\begin{align}
\cR(\cC):=\cL(\cC)-\cH(\cS).
\end{align}

It is well-known that the Huffman encoding algorithm \cite{Huffman52} provides an optimal prefix-free code for a discrete memoryless  source, in the sense that no other code for distribution $\P$ can have a smaller expected length than that of the Huffman code.
A $D$-ary Huffman code is usually represented using a $D$-ary tree $\T$, whose leaves correspond to the source symbols;
The $D$ edges emanating from each intermediate node of $\T$ are labelled with one of the $D$ letters of the alphabet, and the codeword
corresponding to a symbol is the string of labels on the path from the root to the corresponding leaf. 
Huffman's algorithm is a recursive bottom-up construction of $\T$, where at each step the $D$ existing nodes with the smallest  probabilities are merged\footnote{Before that, one has to add a number of dummy source symbols, all with probability zero, so that the number of source symbols $N$ becomes of the form of $k(D-1)+1$, for some integer $k$.}
into a new node, and henceforth represented by an intermediate node in the tree.  Throughout this paper, unless $D$ is explicitly specified, we talk about the binary Huffman codes ($D=2$).

We slightly modify the notation, and write the entropy, average length, and redundancy of a Huffman code in terms of its tree, $\cH(\T)$, $\cL(\T)$, and $\cR(\T)$, respectively. 
It is clear that the redundancy is always non-negative, and easy to show that the redundancy of a Huffman code never exceeds $1$ \cite{Huffman52}.
These bounds on $\cR(\T)$ can be improved if partial knowledge about the source distribution is available.
Gallager \cite{Gallager78}, Johnsen \cite{Johnson80}, Capocelli and Desantis \cite{Cap-DeS91}\cite{CapSan89}, Manstetten \cite{Manstetten92} and Capocelli \emph{et al.} \cite{Capocelli86} improved the upper bound on the redundancy (of binary Huffman codes) when $p_1:=\max_i u_i$, the probability of the most likely source symbol is known. The problem of upper bounding the redundancy in terms of $p_N:=\min_i u_i$, the probability of the least likely source symbol, is addressed in \cite{Cap-DeS91} and \cite{prisco-santis}.   Capocelli \emph{et al.} \cite{Capocelli86} obtained upper bounds on $\cR(\T)$ when both extreme probabilities, $p_1$ and $p_N$, are known. Furthermore,  in \cite{prisco-santis} and \cite{Yeung91} upper bound on the redundancy is derived as a function of probability of the two least likely source symbols, $p_{N-1}$ and $p_N$.

Johnsen \cite{Johnson80} presented a tight lower bound on the redundancy of binary Huffman code in terms of $p_1$ when $p_1\geq0.4$. Subsequently, such lower bounds were generalized for all $p_1$  by Montgomery and Abrahams \cite{Mon-Abr}. Later Goli\'{c} and Obradovi\'{c} \cite{GolObr90} extended Johnsen's result to lower bound the redundancy of  $D$-ary Huffman code in terms of the probability of the most likely symbol. The lower bound on $\cR(\T)$, when only $p_N$ is known, is considered in \cite{Cap-DeS91}. Furthermore, the problem of lower bounding $\cR(\T)$ for a binary code when the two least likely probabilities, $p_{N-1}$ and $p_N$, are known, is discussed in \cite{Cap-DeS91} and \cite{Yeung91}.

Ye and Yeung  raised the problem of bounding the redundancy of  Huffman code when the probability of \emph{one} of the source symbols  (regardless of its order) is known in \cite{YeYeu02}, wherein they presented an upper bound on $\cR(\T)$. In this problem, the assumption is that we a-priori know that the source contains a symbol with a given probability $p$, without knowing about its \emph{rank} in the source distribution, as opposed to
the case when the least or the most likely probability is given\footnote{It is clear that the known probability is in fact the probability of the most likely symbol if $p\geq 0.5$, i.e., $p_1=p$.}. A parametric upper bound for the redundancy is presented in \cite{YeYeu02}, which is \emph{not} tight in general. However, the authors conjectured another upper bound on $\cR(\T)$ in terms of the given probability $p$, which improves the other one. 

In this work, we prove this conjecture with a simple approach and prove that this upper bound is tight. Moreover, we present a tight lower bound on $\cR(\T)$ for a source that contains a symbol with given probability $p$.  We further characterize all possible sets of distribution which achieve this lower bound.  We show that simple extensions of our
results lead to the lower bound on the redundancy when either $p_1$ \cite{Mon-Abr} or $p_N$ \cite{Cap-DeS91} are known. We also extend our proof to the $D$-ary Huffman codes and find the tight lower bound on $\cR(\T)$ when probability of any symbols is known.

The rest of this paper is organized as follows. First, in Section~\ref{sec:notations} we introduce the notation used in the paper and present  lemma which plays a key role this work. Then, we state our main results in Section~\ref{results}. The proof and discussions for the upper bound and lower bound are presented in Section~\ref{sec:up} and Section~\ref{sec:low}, respectively. The lower bound is extended to the $D$-ary Huffman codes in Section~\ref{sec:D}. Finally, we conclude the paper in Section~\ref{sec:con}.

\section{Notations and Preliminaries}\label{sec:notations}
In this section we present some definitions and review some known results that will be useful in the rest of the paper. We start with the definition of multiset \cite{multiset}. 

\begin{definition}
\label{def:multiset}
Let $U$ be a universe set. A \emph{multiset} over $U$ is defined as a pair $\mathbf{F}=\langle U,f\rangle$, where the multiplicity function $f:U\rightarrow \mathds{N} \cup \{0\}$ identifies the number of appearance of each element of the universe in the multiset. 

Operations over sets can be generalized for multisets as follows. 
\begin{itemize}
\item Membership: An $x\in U$ is a member of $\mathbf{F}$,  if $f(x)>0$.
\item Subset: A multiset $\mathbf{F}=\langle U,f\rangle$ is a subset of $\mathbf{G}=\langle U,g\rangle$, and denoted by $\mathbf{F} \subseteq \mathbf{G}$, if $f(x) \leq g(x)$ for any $x\in U$.
\item Union: The union of two multisets $\mathbf{F}=\langle U,f\rangle$ and $\mathbf{G}=\langle U,g\rangle$, denoted by $\mathbf{H}=\mathbf{F} \uplus \mathbf{G}=\langle U, h \rangle$, is a multiset over $U$, with the multiplicity function 
\begin{align*}
h(x)=f(x)+g(x).
\end{align*}
\item Multiset removal: The removal of multiset $\mathbf{G}=\langle U,g\rangle$ from $\mathbf{F}=\langle U,f\rangle$, denoted $\mathbf{H}=\mathbf{F} \ominus \mathbf{G}$, is the multiset $\mathbf{H}=\langle U,h\rangle$ , where 
\begin{align*}
h(x)=\max\{ f(x)-g(x),0\}.
\end{align*}
\end{itemize}
\end{definition}

In the rest of this work, whenever we talk about a probability distribution, it is referred to a multiset $\P$ defined over the universe $U=[0,1]$.

Consider a Huffman tree, with its end nodes (leaves) denoting  the source symbols. In the following, we identify each node (including leaves and intermediate nodes) of the tree by its {\em  probability}; this is defined as the probability of the corresponding symbol for the leaf nodes, and the sum of the probabilities of all the leaf nodes lying in the sub-tree under the node for the intermediate (non-leaf) nodes. In particular, the probability of the \emph{root} is $1$. 

For each intermediate node $u$ in $\T$, denote by $\P_u\subseteq\P$ the multiset of probabilities of the source symbols whose corresponding nodes lie \emph{under} $u$. It is clear that $\sum_{p_i\in\P_u} p_i =u$. Let $\Delta_u$ be the sub-tree of $\T$ under $u$, and denote by
$u^{-1}\ast\Delta_u$ its normalized version, where the probability of each node in $\Delta_u$ is scaled by a factor of $1/u$,
so that the scaled leaf probabilities sum to one. 
Therefore, $u^{-1}\ast\Delta_u$ itself is a Huffman tree for a source with the probability distribution $u^{-1}\P_u$.

\begin{figure}
\begin{center}
	\psfrag{u}[Bc][bc]{$u$}
	\psfrag{d}[Bc][bc]{$\Delta_u$}
	\psfrag{l}[Bc][bc]{$\Lambda_u$}
	\includegraphics[height=5cm]{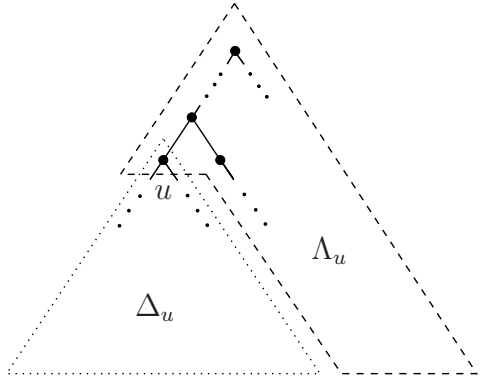}
\end{center}
\caption{Decomposition of a Huffman tree with respect to the intermediate node $u$.}
\label{def}
\end{figure}

Similarly, denote by $\Lambda_u$ the part of the Huffman tree \emph{above} $u$ by collapsing the sub-tree $\Delta_u$ to a single node with probability $u$. It is easy to verify that $\Lambda_u$ is a valid Huffman tree for a source with probability distribution $\P^u := (\P \ominus \P_u ) \uplus \{u\}$. See Figure~\ref{def} for a schematic diagram of the
relationship between $\Delta_u$, $\Lambda_u$,  and the original Huffman tree $\T$.

The following lemma relates the redundancy of a Huffman tree to the redundancies of the sub-trees $(u^{-1}\ast\Delta_u)$ and $\Lambda_u$.

\begin{lemma}\label{lem1}
For any intermediate node $u$ in a Huffman tree $\T$, we have
\begin{equation}
\cR(\T)=\cR(\Lambda_u)+u\cdot \cR(u^{-1}\ast\Delta_u)
\end{equation}
\end{lemma}

\begin{proof} It is well known that the average length of any Huffman code equals the sum of the probabilities on the intermediate nodes (including the root) in the corresponding tree.
Each intermediate node of $\T$ is an intermediate node  either in $\Lambda_u$ or in $(u^{-1}\ast\Delta_u)$, where the probabilities in the latter tree need to be scaled back by a factor of $u$.  Note the node $u$ is not double counted in both trees, because it a  leaf in $\Lambda_u$, and so not counted in the first term. Therefore we have
\begin{eqnarray}\label{L1}
\cL(\T)=\cL(\Lambda_u)+u\,\cL(u^{-1}\ast\Delta_u).
\end{eqnarray}
On the other hand, we can rewrite the entropy of the original tree (source) in terms of the entropy of its associated sub-trees by decomposing the leaf nodes of $\T$ to those of $\Lambda_u$ and $(u^{-1}\ast\Delta_u)$. We get
\begin{align}\label{H1}
\cH(\T)
%&=-\sum_{x\in\leaf(\T)}x\log x\nonumber\\
&=-\sum_{p\in\P} p \log p \nonumber\\
&=-\sum_{p\in\P_u}   p \log p -\sum_{p\in \P\ominus \P_u} p\log p -u\log u + u\log u\nonumber\\
&\stackrel{(a)}{=} -\left[\sum_{p\in\P\ominus \P_u}  p \log p + u\log u\right]
-\left[ \sum_{p\in\P_u}   p \log p- \sum_{p\in\P_u} p \log u\right]  \nonumber\\
&= -\sum_{p\in\P^u}   p \log p - u \sum_{p\in\P_u}  \frac{p}{u} \log \frac{p}{u} \nonumber\\
&= -\sum_{p\in\P^u}   p \log p - u \sum_{q\in u^{-1}\P_u}  q \log q\nonumber\\
%&=-\sum_{x\in\leaf(\Lambda_u)\setminus\{u\}}x\log x-\sum_{y\in\leaf(u^{-1}\ast\Delta_u)}u\,y\log u\,y\nonumber\\
%&=-\Bigg(\sum_{x\in\leaf(\Lambda_u)}x\log x-u\log u\Bigg)-\Bigg(u\sum_{y\in\leaf(u^{-1}\ast\Delta_u)}y\log y+u\log u\Bigg)\nonumber\\
&=\cH(\Lambda_u)+u \cH(u^{-1}\ast\Delta_u)
\end{align}
where in $(a)$ we used the fact that $\sum_{p\in\P_u} p= u$. The desired result will be obtained immediately from \eqref{L1} and \eqref{H1}.%\qed
\end{proof}

\section{Main Results}\label{results}
In this section we state the main results of this paper. The first theorem provides a tight upper bound on the redundancy of  Huffman code for a source containing a symbol with a given probability $p$.  Note that $p$ can be  probability of {\em any} symbol, regardless of the rank of the symbol, through all the results of this work.

\begin{theorem}[Tight Upper Bound on Huffman Redundancy]\label{th_max}
Consider the Huffman code for a source with finite alphabet, which includes a symbol with probability $p$, but is otherwise arbitrary.
The redundancy of this code is upper bounded by
\begin{eqnarray}
R_{\max}(p):=\left\{ \begin{array}{ll}
2-p-\H(p), & \textrm{if $0.5\leq p<1$}\\
1+p-\H(p), & \textrm{if $0\leq p<0.5$}
\end{array}\right.
\end{eqnarray}
where $\H(p):=-p \log p - (1-p) \log(1-p)$ is the binary entropy function.
Furthermore, this bound is tight, so there are sequences of source distributions whose Huffman redundancies converge to $R_{\max}(p)$.\\
\end{theorem}

Our next theorem, presents a lower bound for the redundancy of the Huffman code.

\begin{theorem}[Tight Lower Bound on Huffman Redundancy]\label{th_min}
Consider the Huffman code for a source with finite alphabet, which includes a symbol with probability $p$, but is otherwise arbitrary.
The redundancy of this code is lower bounded by
\begin{equation}\label{Rmin}
R_{\min}(p):=m p-\H(p)-(1-p)\log\Big(1-2^{-m}\Big),
\end{equation}
where $m>0$ takes either of the values $\lfloor-\log p\rfloor$ or $\lceil-\log p\rceil$ which minimizes the expression.
Furthermore, this bound is tight, i.e. there exist source distributions containing a symbol with probability $p$, whose Huffman redundancies equal
$R_{\min}(p)$.\\
\end{theorem}

We extend this result and introduce a lower bound for the redundancy of a $D$-ary Huffman code. 

\begin{theorem}\label{th_min_D}
The redundancy of a $D$-ary Huffman code containing a letter with probability $p$ is tightly lower bounded by
\begin{eqnarray}
R_{\min,D}(p)=mp-\mathscr{H}_D(p)-(1-p)\log_D (1-D^{-m})
\end{eqnarray}
where $m$ is either $\lfloor-\log_Dp\rfloor$ or $\lceil-\log_Dp\rceil$ which minimizes the above expression, and $\mathscr{H}_D(p):=\mathscr{H}(p)/\log(D)$
is the $D$-ary entropy function.\\
\end{theorem}

We prove the above theorems in the following sections. We also discuss derivation of some other known results using the techniques of our proofs. 

\section{Upper Bound}\label{sec:up}
In this section we prove the upper bound introduced in Theorem~\ref{th_max}, and then show that this bound is tight by providing sample probability distributions that can achieve this bound. Before proving this theorem, we shall review some known related results, which will be used in our proof.

Our result improves the following bound obtained in \cite{YeYeu02}, and in fact proves a conjecture for the tightest upper bound
given in the same paper.
\begin{theorem}[Theorem~1 in  \cite{YeYeu02}] \label{not_tight}
Let $p$ be the probability of any source symbol. Then the redundancy of the corresponding Huffman code is upper bounded by
\begin{eqnarray}
R_{\mathrm{ub}}(p):=\left\{ \begin{array}{ll}
2-p-\H(p), & \textrm{if $0.5\leq p<1$}\\
0.5, & \textrm{if $\pi_0< p<0.5$}\\
1+p-\H(p), & \textrm{if $p\leq \pi_0$}
\end{array}\right.
\end{eqnarray}
where $\pi_0\simeq 0.18$ is the smallest root of equation $1+p-\H(p)=0.5$.
\end{theorem}
We will skip the proof of this theorem and refer the interested reader to the original
paper \cite{YeYeu02}.  This upper bound is tight when $p\geq 0.5$ or $p\leq \pi_0\simeq 0.18$, but as also suggested in \cite{YeYeu02}, it is not
tight for the central region $\pi_0< p<0.5$.
Thus, we will only consider this central region in our proof for Theorem~\ref{th_max}, and obtain a tight bound for the
redundancy.

We will also use the following upper bound on the redundancy of a source whose {\em most likely} symbol probability is known.  A more
precise form of this bound is presented in \cite{Manstetten92}, and we refer the interested reader to that work for details and proof.
\begin{theorem}[Extracted from Table~I in \cite{Manstetten92}]\label{up_p1}
Let $p_1$ be the probability of the most likely symbol of a source. Then the Huffman redundancy of this source is upper bounded by the
following function,
\begin{eqnarray}
f(p_1)=\left\{\begin{array}{ll}
2-p_1-\H(p_1) & \textrm{if $0.5 \leq p_1 < 1$}\\
3-5 p_1-\H(2 p_1) & \textrm{if $\pi_1 \leq p_1 < 0.5$}\\
\gamma & \textrm{if $ p_1 \leq \pi_1$}\end{array} \right.
\end{eqnarray}
where $\gamma=R_{\max}(\frac{1}{3})=1+1/3-\H(1/3)\simeq 0.415$, and $\pi_1\simeq 0.491$ is a root of $3-5 x-\H(2 x)=\gamma$.
\end{theorem}

\begin{figure}
\center{
\includegraphics[width=\textwidth]{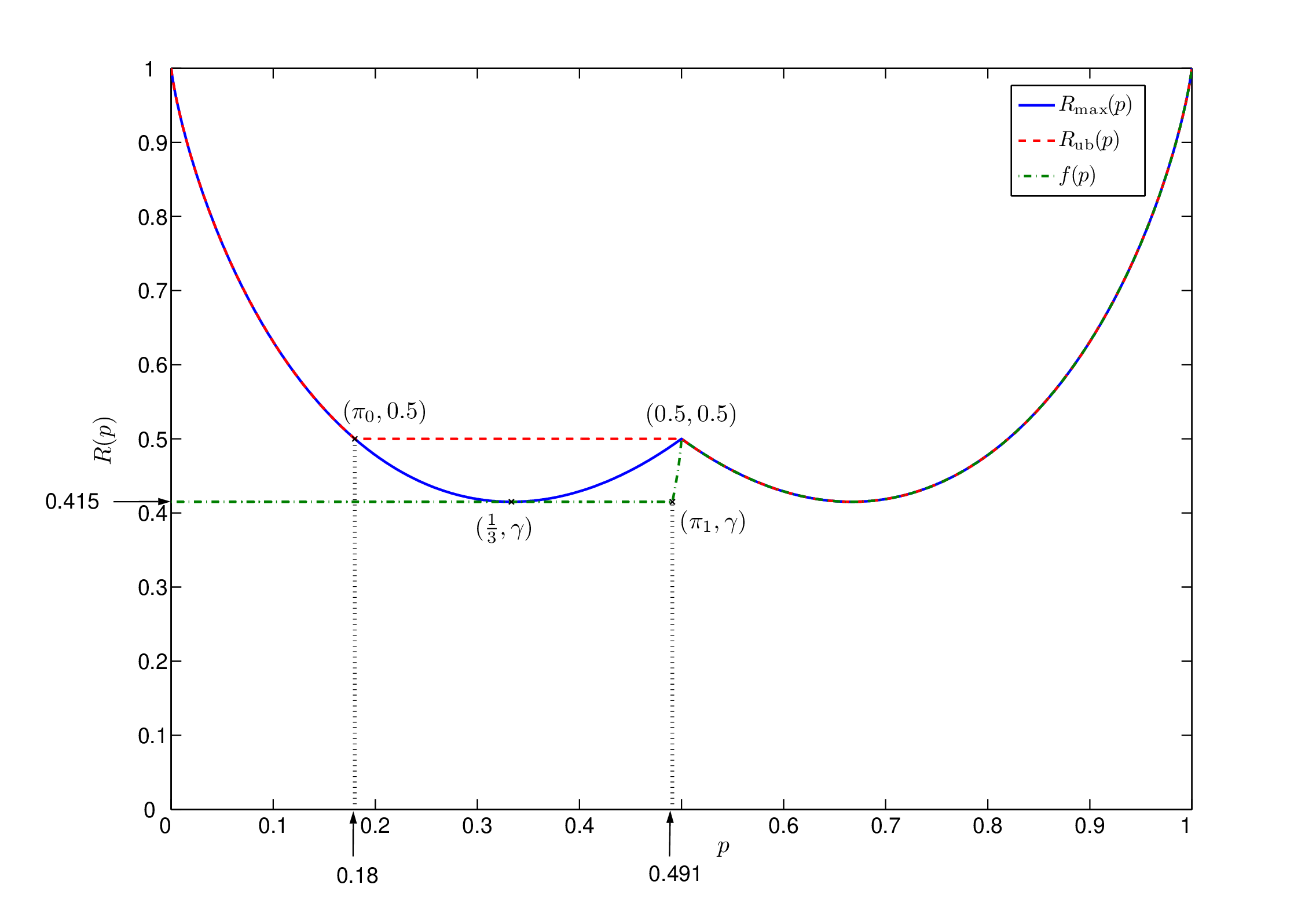}}
\caption{Upper bounds of Theorems~\ref{th_max}, \ref{not_tight} and \ref{up_p1}, on the redundancy of a source containing
a symbol with probability $p$.}
\label{ubounds}
\end{figure}

All the three bounds introduced in Theorems~\ref{th_max},~\ref{not_tight}, and~\ref{up_p1} are illustrated in Figure~\ref{ubounds}.
%In particular this theorem shows that if the probability of the most likely symbol is
%the redundancy of a Huffman code for source with most likely letter of probability $p$,
%is less than or equals to $R_{\max}(\frac{1}{3})=1+\frac{1}{3}-\mathscr{H}(\frac{1}{3})\simeq 0.415$ for $p<0.49$.

%Let $\alpha_1\simeq 0.491$ be the only root of the equation $f(p_1)=R_{\max}(\frac{1}{3})\simeq 0.415$ other than $p_1=\frac{1}{3}$ and
%$p_1=\frac{2}{3}$.  Therefor, for all $p_1\le\alpha_1$, the redundancy of the Huffman code for a source with most likely letter of
%probability $p_1$ is at most $R_{\max}(\frac{1}{3})$.

Finally we will need the following lemma from \cite{Johnson80} in order to prove Theorem~\ref{th_max}. The proof of this lemma is  given in Appendix~\ref{app:lm:l1=1} for completeness.
\begin{lemma}[Theorem~1 in  \cite{Johnson80}] \label{l1=1}
Let $p_1$ be the probability of the most likely letter in a source. If $p_1\geq 0.4$, then the length of the corresponding codeword in the Huffman code is one, i.e., $l(p_1)=1$.
\end{lemma}

We are now ready to prove the upper bound.

\begin{proof}[Proof of Theorem~\ref{th_max}]
As stated before, when $p\leq\pi_0\simeq 0.18$ or $p\geq 0.5$ our bound coincides with that of Theorem~\ref{not_tight}.
It remains to show that for $\pi_0<p<0.5$, the redundancy of a Huffman code for a source which contains a symbol
with probability $p$ does not exceed $R_{\max}(p)$.

We prove this claim using an argument on $p_1$, the probability of the most likely symbol in $\P$.  First note that, if $p_1=p$ is
the most likely symbol, then from Theorem~\ref{up_p1} we simply have $\cR(\T)\leq f(p) \leq R_{\max}(p)$.  

%Suppose then that $p<p_1$. Clearly $p_1\leq 1-p$ since $p$ and $p_1$ are probabilities in the same distribution $\P$.

On the other hand, if $p_1\leq \pi_1$, then from Theorem~\ref{up_p1} the Huffman redundancy of $\P$ cannot exceed
$\gamma=R_{\max}(\frac{1}{3})$. Combining this with the fact that $R_{\max}(\cdot)$ takes its minimum at $p=\frac{1}{3}$,  we have $\cR(\T)\leq f(p_1)=\gamma=R_{\max}(\frac{1}{3}) \leq R_{\max}(p)$. 

So, we can focus our attention to the case where $\pi_0<p<p_1$ and $p_1 > \pi_1$. Note that for $p_1>\pi_1 >0.4$ from Lemma~\ref{l1=1} we  have $l(p_1)=1$. Thus $p$ should appear in the sub-tree under an intermediate node of probability $u:=1-p_1$, i.e. $p\in\P_{u}$, and hence
$q:=\frac{p}{u}\in u^{-1}\ast\Delta_u$. Using Lemma~\ref{lem1}, we can expand the redundancy of the Huffman tree with respect to the intermediate node $u$ which results in 
\begin{align}\label{p_in_p1}
\cR(\T)=  \cR(\Lambda_u) +u \cdot \cR(u^{-1}\ast\Delta_{u})=1-\H(p_1)+u \cdot \cR(u^{-1}\ast\Delta_{u}),
\end{align}
where $1-\H(p_1)$ is the redundancy of the Huffman tree $\Lambda_u$ which only contains two symbols of probabilities $\{p_1,1-p_1\}$.  
We then use Theorem~\ref{not_tight} to upper bound the term $\cR(u^{-1}\ast\Delta_{u})$, which is the redundancy of the Huffman code corresponding to a source contains a symbol of probability $q=p/(1-p_1)$.
Note that $\pi_0<q\leq 1$, since $p$ is assumed to be greater than $\pi_0$.  We consider the following two cases
for the possible values of $q=\frac{p}{1-p_1}$.
\begin{itemize}

\item{\textbf{Case I:} $\pi_0<q\leq 0.5$.}  From Theorem~\ref{not_tight}, we have $\cR((1-p_1)^{-1}\ast\Delta_{(1-p_1)})\leq R_{\mathrm{ub}}(q)\leq\frac{1}{2}$. Replacing this bound in \eqref{p_in_p1}, we get the bound 
\begin{align}
\cR(\T)\leq 1-\mathscr{H}(p_1)+\frac{(1-p_1)}{2},
\end{align}
which can be further bounded by its maximum possible value for $p_1\in(\pi_1,1-\pi_0)$.
Note that the right-hand-side of the above inequality is a convex function of $p_1$ and it takes its maximum value at the boundary point $p_1=1-\pi_0\simeq 0.82$. Then we have
\begin{align}
\cR(\T)&\leq \max_{p_1\in(\pi_1,1-\pi_0)}\left(1-\H(p_1)+\frac{(1-p_1)}{2}\right)\nonumber\\
&=\left(1-\H(\pi_0)+\frac{\pi_0}{2}\right)\simeq 0.410\nonumber\\
&< R_{\max}(p) \hspace{45mm} \forall p\in[0,1]\label{case1}
\end{align}

\item{\textbf{Case II:} $0.5 \leq q\leq 1$.}  Note that $u^{-1}\ast\Delta_u$ is the Huffman tree of source containing a symbol of probability $q\geq 0.5$. Then once again, we can upper bound its redundancy using 
Theorem~\ref{not_tight} as 
\begin{align}
\cR(u^{-1}\ast\Delta_{u}) \leq R_{\mathrm{ub}}(q)=2-q-\H(q).
\label{case2}
\end{align}
Replacing \eqref{case2} in \eqref{p_in_p1} we get
\begin{align}
\cR(\T)&\leq 1-\H(p_1)+(1-p_1)\left(2-q-\H(q)\right)\nonumber\\
&\stackrel{(a)}{\leq} 1-\H(p)+(1-p_1)\left(2-q-\H(q)\right)\nonumber\\
&= 1-\H(p)+p+(1-p_1)\left[2(1- q)-\H(1-q)\right]\nonumber\\
&\stackrel{(b)}{\leq} R_{\max}(p)
\end{align}
where in $(a)$ we used the fact $\H(p_1)\geq \H(p)$ which holds for $p\leq p_1 \leq 1-p$, and the inequality in $(b)$ 
follows from the fact that $(\H(x)-2x)\geq 0$ for $0\leq x<0.5$.
\end{itemize}

Combining \eqref{case1} and \eqref{case2} gives us the desired inequality.
It only remains to show the
tightness of the bound. It is easy to check that the redundancy of a source with distribution
$\P(\epsilon):=\Big((1-\epsilon)(1-p),p,\epsilon(1-p)\Big)$ is
\begin{eqnarray}
\cR(\P(\epsilon))=1+p-\H(p)-(1-p)(\H(\epsilon)-\epsilon),
\end{eqnarray}
for $p\leq (1-\epsilon)(1-p)$, and 
\begin{eqnarray}
\cR(\P(\epsilon))=2-p-\H(p)-(1-p)\H(\epsilon),
\end{eqnarray}
for $p\geq (1-\epsilon)(1-p)$, which tends to $R_{\max}(p)=1+p-\H(p)$ as $\epsilon$ goes to zero. Note that the redundancy is a discontinuous function of the probability distribution at the boundary of this  distribution space. More precisely, at the extreme case where $\epsilon=0$, we have a source with only two symbols, for which the Huffman redundancy is $1-\H(p)$. This completes the proof.
\end{proof}

\section{Lower Bound}\label{sec:low}

In this section we provide the proof of Theorem~\ref{th_min}. We further show that the lower bound is tight, and we identify all distributions which achieve this redundancy. 
%In particular, this theorem shows that the redundancy can be zero only if $p$ is dyadic, i.e $p=2^{-l}$ for some integer $l$.

\begin{proof}[Proof of Theorem~\ref{th_min}]
We first note that, for the purposes of minimizing the redundancy, it suffices to only consider a simple class of probability
distributions for which the corresponding Huffman tree has a \emph{canonical structure}, depicted in Figure~\ref{min_structure}.  To see this, let $u$ be any intermediate node in a Huffman tree $\T$, which
does not contain $p$ in the sub-tree under it, i.e. $p\notin\P_u$.  Then from Lemma~\ref{lem1},
\begin{equation}\label{backbone}
\cR(\T)=\cR(\Lambda_u)+u R(u^{-1}\ast\Delta_u) \geq \cR(\Lambda_u).
\end{equation}
Therefore the redundancy of $\Lambda_u$, the  Huffman tree for the source with probability distribution $\P^u$, which still contains a symbol of probability $p$, does not exceed that of $\T$.  A similar 
argument holds for any sub-tree which does not contain $p$ as a leaf. It is clear that this process converts  $\T$ into the canonical form of Figure~\ref{min_structure}, in which each intermediate node has $p$ in its associated sub-tree, and the redundancy of the canonical form tree is at most the same as that of the original tree.

\begin{figure}
\begin{center}
	\psfrag{x1}[Bc][bc]{$x_1$}
	\psfrag{x2}[Bc][bc]{$x_2$}
	\psfrag{xm1}[Bc][bc]{$x_{m-1}$}
	\psfrag{xm}[Bc][bc]{$x_{m}$}
	\psfrag{p}[Bc][bc]{$p$}
	\includegraphics[height=6cm]{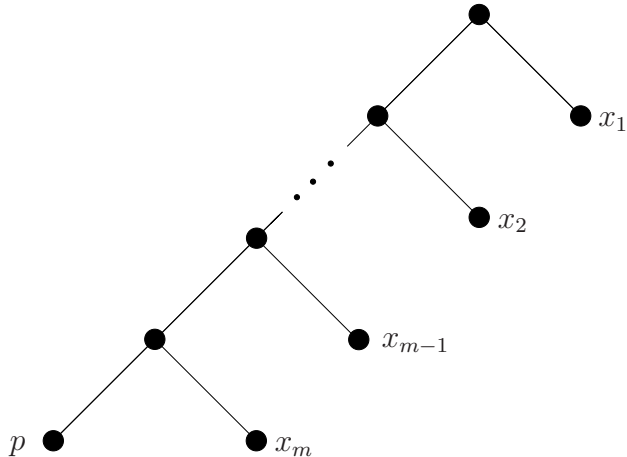}
\end{center}
\caption{Canonical structure for minimum-redundancy Huffman trees that contain a symbol with probability $p$.}
\label{min_structure}
\end{figure}

Suppose then, without loss of generality, that $\T$ is a canonical form Huffman tree of depth $m$, which achieves the minimum redundancy. Let $\P=\{x_1,x_2,\dots,x_m,p\}$ be the corresponding probability distribution. Note that either $p$ or $x_m$ is the probability of the least likely symbol. Define $\alpha_i=x_i/(1-p)$, where $\sum_{i=1}^{m} \alpha_i=1$. The expected length of this code can be written as 
\begin{align}
\cL(\T)&=\sum_{i=1}^m x_i\cdot i+p\cdot m\nonumber\\
&=1+(m-1)p +(1-p)\sum_{i=1}^m (i-1)\alpha_i\nonumber
\end{align}
and the entropy of the source can be expressed in terms of $\alpha_i$'s as
\begin{align}
\cH(\T)&=-\sum_{i=1}^m (1-p)\alpha_i\log[(1-p)\alpha_i]-p\log p\nonumber\\
&=\mathscr{H}(p)+(1-p) H(\alpha_1,\alpha_2,\dots,\alpha_m),\nonumber
\end{align}
where $H(\alpha_1,\alpha_2,\dots,\alpha_m)=-\sum_{i=1}^m \alpha_i \log_2 \alpha_i$ is the entropy of the corresponding distribution. Thus,
\begin{align}
\cR(\T)&=1+(m-1)p-\mathscr{H}(p)+(1-p)\Big[\sum_{i=1}^m (i-1)\alpha_i-H(\alpha_1,\alpha_2,\dots,\alpha_m)\Big]\nonumber\\
&=1+(m-1)p-\mathscr{H}(p)+(1-p)g(\alpha_1,\alpha_2,\dots,\alpha_m)\label{RTexpand}
\end{align}
where $g(\alpha_1,\alpha_2,\dots,\alpha_m):=\sum_{i=1}^m (i-1)\alpha_i-H(\alpha_1,\alpha_2,\dots,\alpha_m)=\sum_{i=1}^m \alpha_i \left(i-1+\log \alpha_i\right)$.
The goal is to find the values of $\alpha_i$ which minimize $g(\cdot)$ (and therefore $\cR(\T)$) for fixed value of $m$. Note first that
the minimizing probability vector $(\alpha_1,\dots,\alpha_m)$ must be an interior point in the probability simplex, since the redundancy function is discontinuous on the boundary of the simplex as mentioned before. More precisely, if
$\alpha_m=0$, one can remove $x_m=0$ from the distribution, ---replacing $m$ with $(m-1)$,--- and reduce the redundancy.  
%Next note that $R(\T)$ depends on the $\alpha_i$'s only through $g(\cdot)$.
Next note that $g(\cdot)$ is a convex function. So, we can use the Karush-Kuhn-Tucker theorem \cite{Boyd} to  find its minimum, subject to the constraints  $\sum_{i=1}^{m}\alpha_i=1$ and $1\geq \alpha_1\geq \alpha_2\geq \cdots \geq \alpha_m\geq 0$. More precisely, we have to solve the following optimization problem
\begin{align}
\left\{
\begin{array}{rl}
\min_{(\alpha_1,\dots,\alpha_m)\in[0,1]^m} &\sum_{i=1}^m \alpha_i \left(i-1+\log \alpha_i\right)\\
\mathrm{subject\  to } &\sum_{i=1}^m \alpha_i-1=0\\
& \alpha_{i+1}-\alpha_{i}\leq 0,\quad i=0,\dots, m,
\end{array}\right.
\label{eq:opt}
\end{align}
where we define $\alpha_0:=1$ and $\alpha_{m+1}:=0$ for consistency. Finding the solution of this optimization problem is more involved and presented in Appendix~\ref{app:opt}. Here we just report the optimum solution which is 
\begin{eqnarray}
\alpha^*_i=\frac{2^{m-i}}{2^m-1}, \qquad i=1,\dots,m. \label{alphas}
\end{eqnarray}
Plugging the optimal values into \eqref{RTexpand} and after straightforward manipulations, we get
\begin{align*}
\cR(\T)= m p-\H(p)-(1-p)\log\Big(1-2^{-m}\Big).
\end{align*}
This is readily seen to be a convex function of $m$.  To minimize, we differentiate with respect to $m$ and set the derivative equal to zero.
\begin{eqnarray*}
\frac{\partial\ \  }{\partial m} \cR(\T)=p-\frac{1-p}{2^m-1}=0,
\end{eqnarray*}
yielding $m=-\log p$.  Since $m$ needs to be an integer and by convexity, one of the two neighboring integers $\lfloor-\log p\rfloor$ or
$\lceil-\log p\rceil$ will give the minimum.
It remains to verify that the $\alpha_i$ values of \eqref{alphas} are consistent with a Huffman tree of the form in Figure~\ref{min_structure}.
A necessary and sufficient condition for this is that $p\leq x_{m-1}=(1-p)\alpha_{m-1}$.  It is then easy to see that the chosen value of $m$
in $\{\lfloor-\log p\rfloor, \lceil-\log p\rceil\}$ which minimizes \eqref{Rmin}, results in an $\alpha_{m-1}$ coefficient which satisfies this
condition.

Finally note that in this proof we have explicitly obtained probability distributions that achieve the minimum bound. This shows the tightness 
of the bound, and completes the proof. 
\end{proof}

\begin{figure}
\center{
\includegraphics[width=\textwidth]{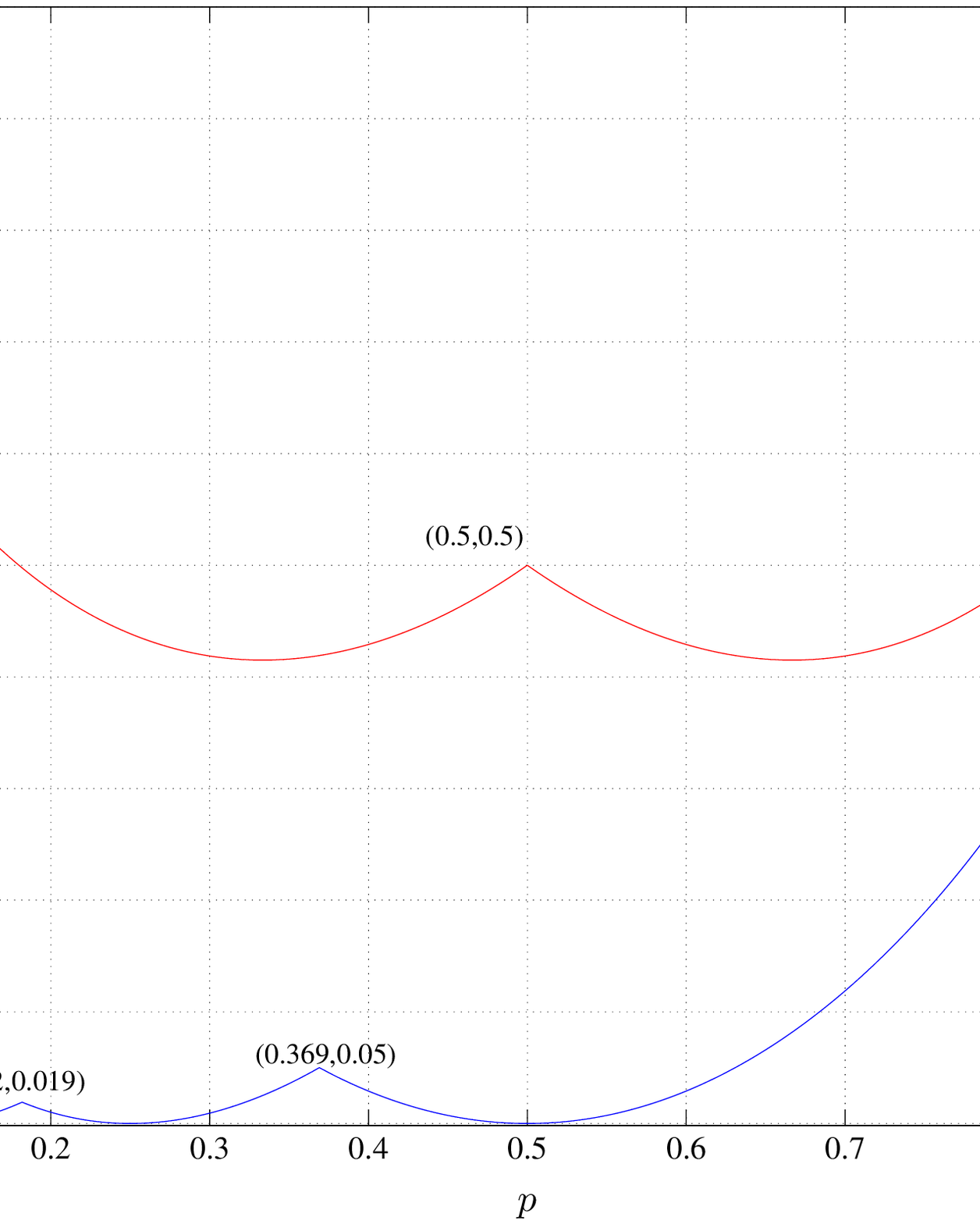}}
\caption{Lower bound and Upper bound on the redundancy of a source containing a symbol with probability $p$.}
\label{simul}
\end{figure}

In the following, we point out a few remarks about Theorem~\ref{th_min}. 

\begin{remark}
The minimizing value of $m$ in Theorem~\ref{th_min} can be precisely expressed as follows.
The optimal $m$ satisfies
\begin{eqnarray}
\beta_{m}\leq p\leq\beta_{m-1}\label{mdef}
\end{eqnarray}
where $\beta_0:=1$ and $\beta_k$ is given by
\begin{eqnarray*}
\beta_k:=\left(1+1/\log\big(1+\frac{1}{2^{k+1}-2}\big)\right)^{-1}.
\end{eqnarray*}
This equation is obtained by equating the values of $\cR(\T)$ from \eqref{Rmin} for two consecutive integers $\lfloor-\log p\rfloor$ and
$\lceil-\log p\rceil$.  It is easy to see that
$\beta_k$ is a descending sequence, converging to $0$ as $k$ tends to infinity, so that for any $p\in (0,1)$ there exists a unique $m$
satisfying \eqref{mdef}.  The first few $\beta_k$'s are $\beta_1=0.369, \beta_2=0.182, \beta_3=0.091, \dots$ and are displayed in
Figure~\ref{simul}.
\end{remark}

\begin{remark}\label{remark-2}
From \eqref{Rmin}, the lower bound $R_{\min}(p)$ can be zero if and only if $p=2^{-m}$ is \emph{dyadic}.  In that
case, from \eqref{alphas} $x_i=(1-p)\alpha_i=2^{-i}$.  Therefore the entire distribution is dyadic.
\end{remark}

\begin{remark} The proof of Theorem~\ref{th_min} essentially describes all the source distributions that contain a symbol
of probability $p$, and achieve the lower bound $R_{\min}(p)$ on the redundancy.  As argued before, the Huffman trees for all such distributions
have a `backbone' of canonical form shown in Figure~\ref{min_structure}, with probabilities that are uniquely determined in the proof of the theorem.  
Any such tree which extends beyond this
unique backbone must satisfy the inequality in \eqref{backbone} with equality, i.e. $\cR({x_i}^{-1}\ast\Delta_{x_i})$ must be zero for all
intermediate $x_i$'s.  From Remark~\ref{remark-2} above, this can happen only if the corresponding distributions for the sub-trees
${x_i}^{-1}\ast\Delta_{x_i}$ are dyadic.  Thus, all the distributions containing a symbols of probability $p$, which achieve the lower
bound $R_{\min}(p)$ can be obtained in the following way: Start with the backbone distribution described in Theorem~\ref{th_min}.  At any time, choose a leaf node other than $p$ and split its probability in half.  It can be shown that each tree during this process is a valid Huffman tree with redundancy $R_{\min}(p)$.
\label{remark-3}
\end{remark}

In the remainder of this section, we extend the results of Theorem~\ref{th_min} to the cases when the given probability $p$ corresponds
to the most, or the least likely symbol. This leads us to two known theorems lower bounding the redundancy for the corresponding cases. 

\subsection{A Lower Bound when the Maximum Probability is Known}
The following theorem extracted from \cite{Mon-Abr} presents a tight lower bound on the redundancy of the Huffman code assigned to a source for which the probability of the most likely symbol is known. We will show in the proof that this theorem is a consequence of Theorem~\ref{th_min}.

\begin{theorem} (Theorem~2 in \cite{Mon-Abr})
A tight lower bound for the Huffman redundancy of a source whose maximum symbol probability is $p_1$ is $R_{\min}(p_1)$, as defined in
Theorem~\ref{th_min}.
\end{theorem}

\begin{proof}
Suppose that $p=p_1$ is the probability of the most likely symbol of a source. It is clear that $R_{\min}(p_1)$ is a lower bound for the redundancy of the corresponding Huffman code. However, it is not clear whether this bound in tight.

In the following we argue that for any $p_1\in(0,1)$, there exist sources whose most likely symbol has probability $p_1$, and its Huffman redundancy achieves $R_{\min}(p_1)$. Let $\mathbf{Q}$ be a distribution containing $p_1$  (not necessarily as the maximum
probability) which achieves the minimum redundancy $R_{\min}(p)$. Then, by the argument of Remark~\ref{remark-3} above, each symbol probability of $\mathbf{Q}$ other than $p_1$ can be successively split into
two halves, without changing the redundancy of the code.  We can repeat this process until $p_1$ becomes the largest value in the distribution.  Therefore, $\mathbf{Q}$ converts to a  desired distribution which contains $p_1$ as the maximum probability, and its Huffman redundancy  equals $R_{\min}(p_1)$.
\end{proof}

\subsection{A Lower Bound when the Minimum Probability is Known}
Next suppose that $p\leq 0.5$ is constrained to be the probability of the least likely symbol of a source.  The next theorem of Capocelli and Santis \cite{Cap-DeS91} characterizes the minimum achievable redundancy of the Huffman code in terms of the probability of the least likely symbol. 

\begin{theorem}\label{th_min_min}(Theorem~2 in \cite{Cap-DeS91})
A tight lower bound for the Huffman redundancy of a source whose minimum symbol probability is $p_N=p$, is given by
\begin{align}
\min \Bigg\{ &p \lfloor\log \frac{1}{p}\rfloor-\H(p)-(1-p)\log\Big(1-2^{-\lfloor \log \frac{1}{p}\rfloor} \Big), \nonumber\\
&  2p \lceil\log \frac{1}{2p}\rceil-\H(2p)-(1-2p)\log\Big(1-2^{-\lceil \log \frac{1}{2p}\rceil}\Big) \Bigg\}.
\label{eq:min_min}
\end{align}
\end{theorem}

\begin{proof}
The argument used in the
proof of Theorem~\ref{th_min} can be extended to this case. In fact, we have to solve a similar optimization problem, except an additional constraint $x_m\geq p$. Writing the KKT conditions \cite{Boyd}, two cases may arise; $(i)$ If the inequality $x_m\geq p$ is strict, then the corresponding coefficient would be zero, and we have exactly the same solution as we had in Theorem~\ref{th_min}. It is easy to see the optimal value for $m$ for which satisfies $x_m\geq p$ is $m=\lfloor -\log p \rfloor$. $(ii)$ If the constraint $x_m\geq p$ becomes tight, then the optimal value for $x_m$ is determined, and one has to solve the system of equations for the remaining $m-1$ variables. In this case, it turns out that the optimal value for $m$ is $m=\lceil -\log 2p \rceil$. The solution of the optimization problem in two cases gives us the bound claimed in the theorem.

The tightness of the bound can be shown by verifying that the two functions in the minimization expression are met by sources with probability distributions $\P_{\min}^{(1)}=\{2^{m-i}(1-p)/(2^m-1): m=\lfloor -\log p \rfloor, i=1,2,\dots,m\} \uplus \{p\}$ and $\P_{\min}^{(2)}=\{2^{m-i}(1-2p)/(2^m-1): m=\lceil -\log 2p \rceil, i=1,2,\dots,m-1\} \uplus \{p,p\}$, respectively. 
\end{proof}

It is worth mentioning that the proof of Theorem \ref{th_min_min} presented here is completely different and simpler than the proof given in the original paper \cite{Cap-DeS91}. The main simplification here is due to the restriction of the search space to the class of canonical Huffman trees, which is a consequence of Lemma~\ref{lem1}.

\begin{remark}
As mentioned before, the bound in Theorem~\ref{th_min_min} coincides with $R_{\min}(p)$ for some regimes of $p$, where $x_m\leq p$. The other function in the minimization expression \eqref{eq:min_min} is precisely $R_{\min}(2p)$, corresponding to the case in Theorem \ref{th_min} when the optimal tree depth is $m=\lceil -\log 2p \rceil$. This, of course, is not a coincidence, since one can easily verify that for a source whose two least probable symbols have probability $p$, i.e. $p_N=p_{N-1}=p$, merging the two corresponding leaves in the Huffman tree results in a tree with a single node of probability $2p$, and with the same redundancy as the original tree. This is because the subtree under the intermediate node $2p$ in the original tree has zero redundancy, since it represents a uniform distribution $(p,p)$. 
\end{remark}

\begin{figure}
\center{
\includegraphics[width=\textwidth]{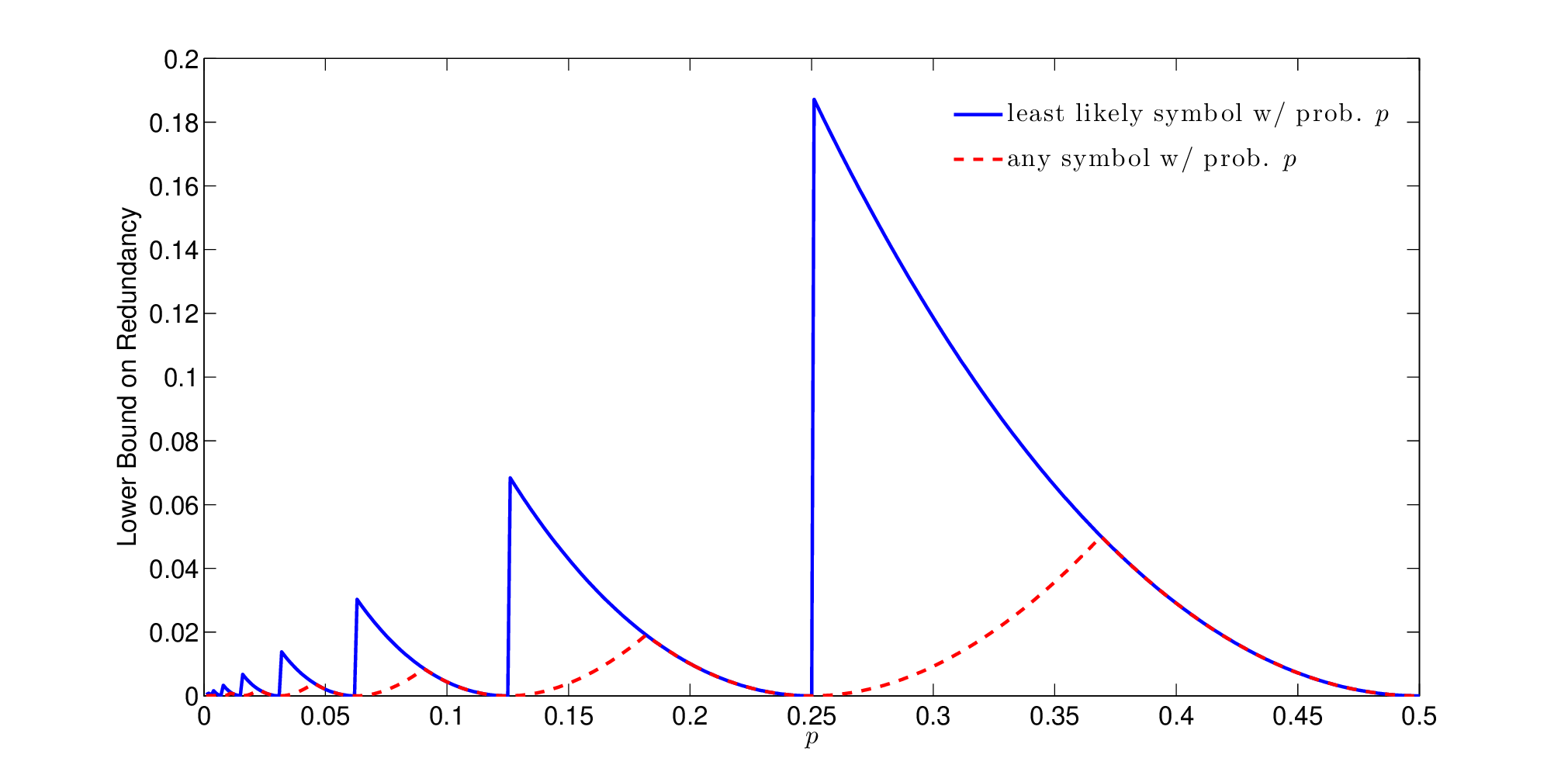}}
\caption{Lower bounds for the Huffman redundancy of  sources containing a symbol with probability $p$ with arbitrary rank, and least likely symbol with probability $p$.}
\label{lowerbounds}
\end{figure}

Figure~\ref{lowerbounds} plots the lower bounds in Theorem~\ref{th_min} and Theorem~\ref{th_min_min} as a function of the fixed probability $p$.

\section{Extension to the $D$-ary Huffman Codes}\label{sec:D}

In this section we present the proof of Theorem~\ref{th_min_D}. Similar to the binary case, the basic idea in this proof is restriction to ($D$-ary) canonical Huffman trees, and solving the optimization problem to find the optimal values for the symbol probabilities. We need the following lemma to simplify the proof. We refer to Appendix~\ref{app:lm:equal} for its proof. 

\begin{figure}
\begin{center}
 	\psfrag{x_12}[Bc][Bc]{$x_{1,2}$}
	\psfrag{x_1D}[Bc][Bc]{$x_{1,D}$}
	\psfrag{x_22}[Bc][Bc]{$x_{2,2}$}
	\psfrag{x_2D}[Bc][Bc]{$x_{2,D}$}
	\psfrag{x_m-12}[Bc][Bc]{$x_{m-1,2}$}
	\psfrag{x_m-1D}[Bc][Bc]{$x_{m-1,D}$}
	\psfrag{p}[Bc][Bc]{$p$}
	\psfrag{x_m2}[Bc][Bc]{$x_{m,2}$}
	\psfrag{x_mD}[Bc][Bc]{$x_{m,D}$}
\includegraphics[width=9cm]{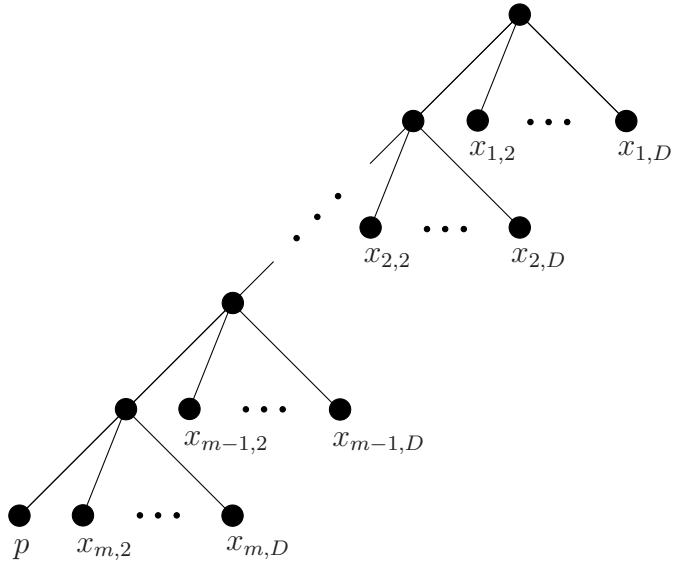}
\end{center}
\caption{Canonical structure for minimum-redundancy $D$-ary Huffman trees that contain a symbol with probability $p$.}
\label{D-ary}
\end{figure}

\begin{lemma}\label{lm:equal}
The minimum-redundancy $D$-ary Huffman tree for a source has a canonical structure as shown in Figure~\ref{D-ary}. Moreover, all the leaves  at the same depth (except possibly $p$) have the same probability, i.e., $x_{l,i}=x_{l,j}$ for all $l=1,\dots,m$ and  $i,j=2,\dots,D$, where $x_{l,i}$ is the $i$-th node in depth $l$ of the tree. 
\end{lemma}

\begin{proof}[Proof of Theorem~\ref{th_min_D}]
Using Lemma~\ref{lm:equal}, it suffices to only consider Huffman trees of $D$-ary canonical form to find the minimum redundancy tree. Similar to the binary case,  we have to solve the optimization problem for $\bar{x}_1,\dots,\bar{x}_m$, where $x_{l,i}=\bar{x}_l$ for $i=2,\dots,D$ and $l=1,\dots,m$. The details of the solution are very similar to the binary case and we skip them for brevity. It turns out that the optimal solutions are
\begin{align}
\bar{x}_l=\frac{(1-p) D^{m-l}} {D^m-1}, \qquad l=1,2,\dots,m.
\end{align}
Replacing these values in the redundancy expression, we get
\begin{eqnarray}
R_{\min,D}=mp-\mathscr{H}_D(p)-(1-p)\log_D (1-D^{-m}),
\end{eqnarray}
which is minimized at $m^\ast=-\log_D p$. The convexity of the function with respect to $m$ implies that the optimal depth belongs to the set $\{ \lfloor-\log_Dp\rfloor, \lceil-\log_Dp\rceil  \}$. This completes the proof. 
\end{proof}

\section{Conclusion}\label{sec:con}
We introduced a lemma to expand the average length, entropy, and the redundancy of a Huffman tree with respect to any of its intermediate nodes. This method simplifies analyzing the redundancy behavior of the tree. In particular, we used this method to obtain tight upper and lower bounds for the Huffman tree associated to a source containing a symbol of probability $p$, without any further assumption on the rank of the symbol.  The upper bound proves a conjecture of \cite{YeYeu02}. Our lower bound extends and completes several earlier partial results.

We have further discussed the explicit form of the distributions that achieve each of these bounds.  Our arguments can be extended to the case of the
$D$-ary Huffman codes, to obtain a lower bound for its redundancy in terms of any given source symbol probability.

\appendices

\section{Proof of Lemma~\ref{l1=1}}
\label{app:lm:l1=1}
We first note that as long as $p_1>\frac{1}{3}$, the length $l(p_1)$ cannot be larger than $2$; otherwise, there would be at
least two independent intermediate nodes $x$ and $y$ on the Huffman tree  (in the branch does not contain $p_1$) with lengths less than $l(p_1)$.  Therefore $x$ and $y$ both would have
probabilities at least as large as $p_1$; but this is a contradiction since $p_1+x+y\geq 3 p_1$ cannot exceed $1$.

\begin{figure}
\begin{center}
 	\psfrag{p_1}[Bc][Bc]{$p_1$}
	\psfrag{u}[Bc][Bc]{$u$}
	\psfrag{v1}[Bc][Bc]{$v_1$}
	\psfrag{v2}[Bc][Bc]{$v_2$}
\includegraphics[width=9cm]{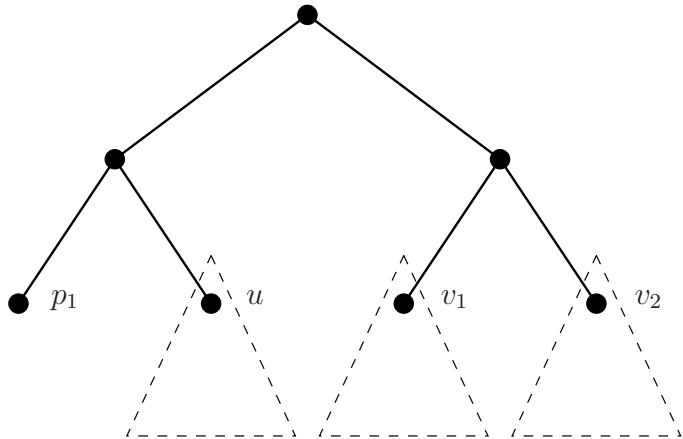}
\end{center}
\caption{A Huffman tree with $l(p_1)=2$.}\label{depth2}
\end{figure}

Suppose next that $l(p_1)=2$.  Then there can be no codeword of length $1$ in the code, since $p_1$ is the largest probability.
Therefore, the corresponding tree has a structure as in Figure~\ref{depth2}.

Note first that $v_1+v_2\geq p_1$ since $l((v_1+v_2))=1<2=l(p_1)$. Moreover, $u\geq \max(v_1,v_2)\geq \frac{1}{2}(u_1+u_2)$ because it is merged after merging $v_1$ and $v_2$. This yields to 
\[
1=p_1+u+v_1+v_2 \geq p_1+ \frac{1}{2}p_1 +p_1=\frac{5}{2} p_1 >1,
\]
which is a contradiction with $p\geq 0.4$. \qed
%We will next show that $u\ge q$.  Clearly, if $q$ lies in the sub-tree under $u$, the assertion follows.  Suppose then that $q$ lies in the sub-tree under $v_1$.  Note next that, by the construction of the Huffman tree, either $p_1, u \geq v_1, v_2$, or $p_1, u \leq v_1, v_2$.  The second alternative cannot happen, since then $1\geq p_1+v_1+v_2\geq 3 p_1\geq 3\pi_1\simeq 1.47 $ would be a contradiction.  Therefore $u\geq v_1\geq q$.

%Combining these relationships we get
%\begin{equation*}
%\pi_1\leq p_1\leq v_1+v_2=1-p_1-u\leq 1-p_1-q\leq 1-\pi_1-\pi_0,
%\end{equation*}
%which is a contradiction, since $\pi_1\simeq 0.491$ and $1-\pi_0-\pi_1\simeq 0.329$.

\section{Optimization Problem in \eqref{eq:opt}}
\label{app:opt}
We first define the Lagrangian
\begin{align}
L(\underline{\alpha}, \lambda, \underline{\mu})=g(\underline{\alpha})+ \lambda \left(\sum_{i=1}^m \alpha_i-1\right) +\sum_{j=0}^m \mu_j (\alpha_{j+1}-\alpha_j),
\end{align}
in which $\underline{\alpha}=(\alpha_1,\dots,\alpha_m)$. Moreover, $\underline{\mu}=(\mu_0,\dots,\mu_m)$ and $\lambda$ are the optimization parameters in the dual program. It is well known that for the optimal values $(\underline{\alpha}^*,\lambda^*, \underline{\mu}^*)$ we have 
\begin{align}
\frac{\partial }{\partial \alpha_i} L(\underline{\alpha}, \lambda^*, \underline{\mu}^*)\Big|_{\underline{\alpha}=\underline{\alpha}^*}  %\left[g(\underline{\alpha}^*)+ \lambda^* \left(\sum_{i=1}^m \alpha^*_i-1\right) +\sum_{j=0}^m \mu_j^* (\alpha_{j+1}^*-\alpha_j^*)\right]\nonumber\\
= (i-1) + \log \alpha_i^* +\frac{1}{\ln 2} + \lambda^* +\mu^*_{i-1}-\mu^*_{i}=0
\label{eq:cond-1}
\end{align}
for $i=1,2,\dots,m$. From the KKT conditions \cite[Section~5.5.3]{Boyd} we have $\mu_j^*\geq 0$ and $\mu_j^* (\alpha_{j+1}^*-\alpha_j^*)=0$. Let $\mu^*_k=\max_{j} \mu_j^*$ be the maximum value among all optimum $\mu_j$'s variables. If $\mu_k^*>0$, from the KKT conditions we get $\alpha_{k+1}^*=\alpha_k^*$. Therefore, from \eqref{eq:cond-1} for $i=k$ and $i=k+1$ we have 
\begin{align}
k-1 + \log \alpha_k^* +\frac{1}{\ln 2} + \lambda^* +\mu^*_{k-1}-\mu^*_{k}&=
k + \log \alpha_{k+1}^* +\frac{1}{\ln 2} + \lambda^* +\mu^*_{k}-\mu^*_{k+1}&=0
\end{align}
which yields
\begin{align}
\mu^*_{k-1} +\mu^*_{k+1}=  2\mu^*_k+1.
\end{align}
Therefore, we have either $\mu^*_{k-1}>\mu_k^*$ or $\mu_{k+1}^*>\mu_k^*$ which are both in contradiction with maximality of $\mu_k^*$. This implies  $\mu^*_i=0$ for  $i=0,\dots,m$. Hence, 
\begin{align*}
i-1 + \log \alpha_i^* +\frac{1}{\ln 2} + \lambda^* =0,
\end{align*}
which implies $\alpha_i^*=K 2^{-i}$ with $K=2^{1-\frac{1}{\ln 2} -\lambda^*}$. Replacing this in $\sum_{i=1}^m \alpha_i^*=1$, we get 
\[
K \sum_{i=1}^m 2^{-i} = K\frac{2^m-1}{2^m}=1,
\]
which results in $K=2^m/(2^m-1)$, and equivalently, 
\begin{align*}
\hspace{52mm}\alpha_i^*=\frac{2^{m-i}}{2^m-1},\qquad i=1,\dots,m. \hspace{52mm}  \blacksquare
\end{align*}

\section{Proof of Lemma~\ref{lm:equal}}
\label{app:lm:equal}
We prove the claim by contradiction. Let $\T$ be a Huffman tree for the source with minimum redundancy, in which $x_{l,i}\neq x_{l,j}$ for some depth $l$, and leave numbers $i$ and $j$. Consider the a new source with corresponding Huffman tree $\T'$ similar to $\T$, except the leaves in depth $l$ are all replaced by $\bar{x}_l:=\frac{1}{D-1}\sum_k x_{l,k}$, the average probability of the corresponding leaves on $\T$. 

Note first that since the average is between the minimum and maximum probabilities in that level, the new tree is still consistent with Huffman tree structure, i.e. $x_{l+1,t}\leq \min_k x_{l,k}<\bar{x}_l<\max_k x_{l,k}\leq x_{l-1,t}$, for all $t=1,\dots,D$.
Next note that the average length of the tree remains fixed after this process, $\cL(\T')=\cL(\T)$.
Finally, expanding the entropy of both sources with respect to the intermediate node $u:=x_{l-1,1}=(D-1)\bar{x}_l$ (the father of the nodes in depth $D$), we have 
\begin{align}
\cH(\T')&=\cH(\Lambda'_{u})+u \cH(u^{-1} \ast \Delta'_u)\nonumber\\
&=\cH(\Lambda'_{u})+u H_D\left( \frac{1}{D-1},\dots, \frac{1}{D-1}\right)\nonumber\\
&\stackrel{(a)}{>} \cH(\Lambda'_{u}) + u H_D\left( \frac{x_{l,2}}{(D-1) \bar{x}_l}, \dots, \frac{ x_{l,D} }{ (D-1)\bar{x}_l } \right)\nonumber\\
&=\cH(\Lambda_{u})+u \cH(u^{-1} \ast \Delta_u)\nonumber\\
&=\cH(\T),
\end{align}
where $H_D(\cdot)$ is the base-$D$ entropy function  which is uniquely maximized
with the uniform distribution as used in $(a)$. Therefore the entropy is maximized, ---and the redundancy minimized,--- with the proposed replacement 
$x_{i,l}\leftarrow\bar{x}_l$, i.e., $\cR(\T')<\cR(\T)$ which is in contradiction with optimality assumption of $\T$.\qed

\bibliographystyle{IEEEtran}

%\IEEEbiographynophoto
\begin{IEEEbiographynophoto}{Soheil Mohajer}
received the B.Sc. degree in electrical engineering from the Sharif University of Technology, Tehran, Iran, in 2004, and
the M.Sc. and Ph.D. degrees in communication systems both from Ecole Polytechnique F\'ed\'erale de Lausanne (EPFL),
Lausanne, Switzerland, in 2005 and 2010, respectively. He was later a post-doctoral research associate in Princeton  University.

Dr. Mohajer has been a post-doctoral researcher at the University of California at Berkeley, since October 2011. 
His research interests include multi-user information theory, statistical machine learning, and bioinformatics. 
\end{IEEEbiographynophoto}

\begin{IEEEbiographynophoto}{Payam Pakzad }
received the B.Sc. in Electrical Engineering and Applied Mathematics
from Caltech, and M.Sc. and Ph.D. in the department of Electrical Engineering
and Computer Science from UC Berkeley.  He was later a post-doctoral
researcher at EPFL.

Dr. Pakzad is currently a researcher with Qualcomm Research Lab in Silicon
Valley.  His research interests include coding theory, machine learning
and statistical signal processing.

\end{IEEEbiographynophoto}

\begin{IEEEbiographynophoto}{Ali Kakhbod}
received the B.Sc. degree in electrical engineering and the B.Sc. degree in mathematics from the Isfahan University Technology, Isfahan, Iran, in 2006 and 2007, respectively, and the M.Sc. degree in electrical engineering and the M.Sc. degree in mathematics from the University of Michigan, Ann Arbor,
in 2009 and 2010, respectively, where he is currently pursuing the Ph.D. degree.

His research interests are in information economics, information theory, stochastic control, and decentralized systems.
\end{IEEEbiographynophoto}

\end{document}